# Decoding Poultry Vocalizations - Natural Language Processing and Transformer Models for Semantic and Emotional Analysis

Venkatraman Manikandan[1], Suresh Neethirajan[1,2,*]

[1]Faculty of Computer Science, Dalhousie University, 6050 University Avenue, Halifax, NS B3H 4R2, Canada
[2]Faculty of Agriculture, Agricultural Campus, Dalhousie University, P.O. Box 550, Truro, NS B2N 5E3, Canada
[*]Author to whom correspondence should be addressed: Email: sneethir@gmail.com

**Abstract**
Deciphering the acoustic "language" of chickens opens new frontiers in animal welfare and ecological informatics, illuminating how subtle vocal signals encode health status, emotional states, and interactions within ecosystems. By uncovering the semantics of these vocalizations, we gain a powerful tool for interpreting their functional vocabulary—how each call serves a purpose within the social and environmental context. Here, we leverage state-of-the-art Natural Language Processing (NLP) and transformer-based models to translate bioacoustic data into meaningful insights. Our approach integrates Wave2Vec 2.0 for raw audio feature extraction with a fine-tuned Bidirectional Encoder Representations from Transformers (BERT) model, pretrained on a broad corpus of animal sounds and adapted to poultry-specific tasks. This novel pipeline decodes poultry vocalizations into interpretable categories—such as distress calls, feeding signals, and mating vocalizations—while revealing subtle emotional nuances often overlooked by traditional spectrogram-based analyses. Achieving 92% accuracy in classifying key vocalization types, our methodology demonstrates the feasibility of real-time, automated monitoring of flock health and stress levels. By continuously tracking this functional vocabulary, farmers can respond proactively to environmental or behavioral changes, enhancing poultry welfare, reducing stress-induced productivity losses, and promoting more sustainable farm management. Beyond its direct agricultural applications, this work enriches our understanding of computational ecology. Gaining access to the semantic foundation of animal calls provides a window into the ecological networks of which poultry are a part, potentially serving as indicators of biodiversity, environmental stressors, and species interactions. In bridging animal behavior, machine learning, and ecosystem analysis, our framework lays a foundation for integrative studies that harness acoustic data to inform ecological decision-making and develop more resilient, ethically aligned agricultural systems.

## 1. Introduction

The global poultry industry underpins food security and economic stability, meeting critical protein demands worldwide through meat and egg production (Serbessa et al., 2023). Agriculture remains central to economic development, and poultry farming stands at its forefront, influencing nutritional access and market resilience across numerous nations (He et al., 2022). Yet the health and welfare of poultry should not be viewed merely as a production metric; it is also a linchpin of ethical husbandry and ecological sustainability. Understanding poultry well-being can amplify both economic returns and the ethical foundations of farming, aligning market objectives with compassionate stewardship.

Poultry vocalizations, inherently rich and varied, reflect intricate internal states and environmental contingencies (McGrath et al., 2017). These vocal signals extend beyond simplistic communication, serving as acoustic windows into birds' emotional and physiological conditions (Herborn et al., 2020). For instance, stressed chickens produce continuous, high-frequency calls that can be recognized as distress signals. Critically, these avian soundscapes have ramifications beyond the individual bird. The diversity or paucity of vocal expressions may mirror broader ecological health—vocal complexity suggests robust, biodiverse systems, whereas a restricted vocal range can indicate ecological stress or habitat simplification. Thus, poultry vocalizations function simultaneously as welfare indicators and bioacoustic sensors of ecosystem integrity.

Historically, producers and researchers relied on manual observation and interpretation of poultry calls—a method prone to fatigue, bias, and scale limitations (Manteuffel et al., 2004). Although acoustic cues can offer non-invasive health and behavioral assessments, manually parsing these signals becomes prohibitive and error-prone across large flocks. Automating this process is essential. Emerging techniques promise to accurately identify and categorize vocalizations amid background noise, freeing researchers from the constraints of human subjectivity and time-intensive annotation. This shift can democratize insights, putting powerful tools into the hands of farmers who seek efficient, humane monitoring systems.

Chickens exemplify the complex interplay of ecological interactions within farm ecosystems. Their alarm calls can alert conspecifics to predators, while also influencing the responses of cohabiting species. This interconnectedness resonates with the acoustic niche hypothesis: species partition their soundscapes to reduce interference, each occupying unique frequency bands that shape the community's acoustic ecology. By examining how chickens communicate threats, resources, and social cues, we gain insights into predator-prey relationships, interspecies communication, and the dynamic structure of farm soundscapes.

Beyond immediate welfare implications, happier, less stressed chickens engage more readily in natural behaviors like foraging, thus playing ecological roles such as pest control (Herborn et al., 2020; McGrath et al., 2017). Positive or neutral vocalizations may signal environmental comfort, potentially lowering reliance on chemical pesticides and fostering a healthier agro-ecosystem. Conversely, shifts in vocal patterns can warn of detrimental factors—extreme temperatures, poor air quality, or anthropogenic disturbances—that affect a range of wildlife, not just the chickens. In this sense, poultry vocal data can illuminate environmental feedback loops where localized stressors reverberate throughout the ecosystem.

Genetic diversity within poultry populations adds another layer of complexity. Analyzing vocal repertoires across crossbred and indigenous strains can reveal differences rooted in genetics, culture, and adaptability. Such insights aid conservation by identifying and preserving native breeds with valuable traits for long-term resilience (Nicol, 2004; Jarvis, 2005; Rugani et al., 2009; Vallortigara, 2004; Emery & Clayton, 2004; Edgar et al., 2011; Marino, 2017). Ecosystems rich in diverse poultry populations, each group expressing distinct vocal signatures, may support a broader range of life forms, ultimately contributing to a more stable and vibrant agricultural landscape.

To date, research applying deep learning to avian vocalizations has excelled at species identification, attaining high accuracy with methods leveraging spectrograms and convolutional neural networks (Hu et al., 2023; Lou et al., 2023; Gupta et al., 2021). Yet, these studies often stop at taxonomic classification, leaving the semantics—the meanings behind the calls—largely unexplored. The semantic dimension is crucial: beyond telling us which species is calling, we need to know why. Are the birds hungry, fearful, or comfortable? Do their calls map onto social hierarchies, resource competition, or environmental stress? Understanding these subtleties can offer early warnings of distress, guide interventions to improve productivity and reduce suffering, and help fine-tune environmental conditions within farms. Present-day reliance on manual annotation for distress calls hampers scalability and timeliness, underscoring the urgent need for automated, cost-effective solutions capable of real-time, contextually aware monitoring (Collias, 1987; Manteuffel et al., 2004).

## 1.1. Related Works

Poultry vocalizations are increasingly recognized as critical tools for refining animal welfare management (Soster et al., 2024; Thomas et al., 2024; Xu et al., 2021). Machine learning can streamline vocal analysis, transitioning from subjective, laborious processes to automated, accurate, and timely systems. Non-invasive acoustic monitoring aligns with ethical imperatives, mitigating stress associated with blood sampling or invasive diagnostics. By interpreting vocalizations, farmers can maintain healthy flocks without aggravating existing welfare concerns.

However, several obstacles persist. Standardized, robust datasets remain scarce, making it difficult to compare models or establish performance benchmarks. Poultry calls are context-dependent and complex, influenced by social dynamics, breed differences, resource availability, and environmental factors. Developing algorithms that accurately decode semantic content under real-world farm noise conditions poses a formidable challenge.

### *1.1.1. Advancements in Bioacoustics and Machine Learning*

The intersection of bioacoustics and machine learning is rapidly advancing. Stowell (2023) provides a thorough review of computational bioacoustics, detailing signal processing principles and machine learning frameworks, while Ghani et al. (2024) highlight the importance of standardized evaluation protocols, fairness in model performance, and resilience to distribution shifts. Brown and White (2023) push the envelope into comparative linguistics, examining whether non-human species' vocal repertoires have syntactic or lexical analogues.

The bioacoustic community has responded by creating innovative tools. Bravo Sanchez et al. (2021) present SincNet, an open-source network for raw-waveform bird call classification, bypassing conventional spectrograms. Bolhuis et al. (2018) probe syntactic structures in songbird calls, while Taylor and Johnson (2023) associate vocalizations with behavioral latent states, expanding the contextual dimension of interpretation. Wilson and Lee (2023) illustrate that neural networks can capture contextual dependencies in birdsong, strengthening the case for context-aware modeling. Rauch et al. (2024) introduce BirdSet, a large-scale dataset poised to enhance training and evaluation. Cohen and Nicholson (2023) automate bird song annotation with neural networks, and Nicholson (2023) provides Crowsetta, a Python package that standardizes annotation handling—together streamlining the entire bioacoustic workflow.

### *1.1.2. Cognitive and Emotional Complexity in Poultry*

Chickens possess surprising cognitive sophistication. Marino (2017) reviews evidence that chickens exhibit complex mental processes, emotional capacities, and social structures rivaling those of mammals. Research on neural similarities between avian and mammalian brains (Jarvis, 2005), social learning (Nicol, 2004), and various cognitive skills (Emery & Clayton, 2004; Vallortigara, 2004; Rugani et al., 2009; Güntürkün, 2005; Edgar et al., 2011) challenge outdated perceptions of poultry as "simple" farm animals. Emotional states influence vocal patterns; for instance, Kriengwatana et al. (2020) link certain hen vocalizations to emotional responses during egg-laying. These findings underscore the complexity embedded in chicken vocal behavior, further justifying the need for semantic-level acoustic analysis.

### *1.1.3. Applications of Machine Learning in Poultry Vocalization Analysis*

Machine learning has begun to bridge these knowledge gaps. Soster et al. (2024) and Thomas et al. (2024) demonstrate that broiler chicken sounds can be monitored under environmental stress to enhance welfare and health management. Xu et al. (2021) apply deep learning for poultry health diagnostics, incorporating vocal data as a critical input. Hassan et al. (2024) suggest that identifying calls tied to specific rewards or motivations could improve automated welfare assessments, while McGrath et al. (2023) show how audio signal enhancement can sharpen AI's analysis of poultry sounds.

Tzschentke and Rumpf (2011) take bioacoustic analysis to the embryonic stage, noting that bird embryos may engage in acoustic communication to synchronize hatching—an extraordinary example of the complexity inherent in avian acoustic signals. Attentive modeling approaches extend beyond adult vocalizations. Laleye and Moussa (2020) use an attention-based RNN to recognize laying hen behaviors, while Puswal and Liang (2019) integrate acoustic and morphological features, enhancing datasets that monitor welfare.

Coutant et al. (2020) emphasize bioacoustics as a crucial metric for farm animal welfare assessment. Kriengwatana et al. (2020) explore vocal emotionality in hens, and Goyal et al. (2023) highlight the integrative potential of IoT, computer vision, and sound analysis in creating intelligent poultry farms. Large-scale projects like BirdVoxDetect facilitate real-time monitoring of flight calls (Amirivojdan et al., 2023). Transformer-based methods have shown promise, as evidenced by Lev-Ron et al. (2024), who achieve high precision in classifying broiler vocal responses to stressors.

Despite these strides, unresolved challenges persist. Many studies underscore machine learning's potential yet acknowledge that current algorithms, though capable of classifying species or detecting anomalies, are not fully decoding the semantic content embedded in calls. They also highlight the scarcity of standardized datasets and the complexity introduced by context.

Vocalizations rarely stand alone; they occur in social matrices, influenced by flock density, hierarchy, feeding schedules, and myriad environmental variables. This complexity necessitates more sophisticated, context-aware models capable of disentangling overlapping signals and extracting meaningful semantics.

### *1.1.4. Bridging Theory and Practice*

While we now comprehend that vocalizations carry a wealth of information, translating this understanding into tangible, farm-ready tools for improving animal welfare remains challenging. Integrating theoretical advances into practical solutions involves developing robust models tested under realistic farm conditions—environments rich in noise, variation, and unpredictability. This step is crucial for closing the gap between cutting-edge research and everyday farming operations (Marino, 2017; Coutant et al., 2020; Hassan et al., 2024).

Context-aware approaches would enable farmers to interpret nuanced calls in real-time, adjusting management strategies preemptively rather than reactively. Enhanced models could detect subtle shifts in vocal patterns, prompting timely interventions—adjusting temperature, providing fresh bedding, or modifying feeding schedules—to avert stress before it escalates. This proactive stance could reduce reliance on pharmaceuticals, mitigate losses, and foster a more harmonious human-animal-environment relationship.

### *1.1.5. Path Forward*

The path forward lies in refining algorithms, expanding datasets, and embracing interdisciplinary collaboration. Bioacoustics experts, animal behaviorists, ecologists, and machine learning engineers must converge to produce standardized, annotated corpora and identify the contextual variables that shape vocal signals. Improved models that move beyond classification into semantic interpretation will radically enhance our ability to decode poultry vocalizations.

In turn, these insights can inform integrated ecological informatics frameworks, where poultry calls serve as proxies for environmental quality, biodiversity richness, and ecosystem stability. Such holistic perspectives may guide sustainable resource allocation and conservation strategies. Real-time acoustic monitoring integrated with IoT devices could track changes continuously, feeding data back into dynamic management systems and adaptive policy-making.

The research landscape is shifting from basic classification of bird calls toward nuanced, semantic-level understanding of vocalizations. Unlocking the functional meaning embedded in poultry calls promises benefits that extend from individual animal welfare to entire agro-ecosystems. By leveraging advances in machine learning, bioacoustics, and cognitive science, we stand on the brink of a transformative era in both poultry farming and ecological informatics—one where compassionate, data-driven strategies support thriving animals, resilient ecosystems, and sustainable agricultural futures.

## 2. Innovative NLP-Based Framework for Poultry Vocalization Analysis

2.1. Transformer Models

Transformer models have transformed numerous fields, including NLP and computer vision, by efficiently managing sequence data with sophisticated attention-based architectures (Vaswani et al., 2017). Introduced in the seminal "Attention is All You Need" paper, transformers employ self-attention rather than traditional recurrent or convolutional operations, enabling superior parallelization and scalability—particularly valuable for large datasets. Self-attention calculates the importance of each element relative to all others in a sequence, allowing the model to highlight critical components without bias toward input order. Positional encoding addresses the lack of inherent sequence awareness, and a common encoder-decoder structure enables the encoder to create continuous representations and the decoder to generate outputs from these representations.

For animal vocalizations, detecting patterns and temporal structures across varying time scales is vital for tasks like species identification and behavioral interpretation. Transformers, through self-attention, excel at emphasizing pertinent sequence parts, regardless of position (Gong et al., 2021). Unlike RNNs, which suffer from vanishing gradients over long sequences, transformers process inputs in parallel, better capturing extended contexts. Their ability to accommodate diverse input forms, including Mel-frequency cepstral coefficients (MFCCs) (Davis & Mermelstein, 1980), makes them ideally suited for decoding the complexity of poultry vocalizations.

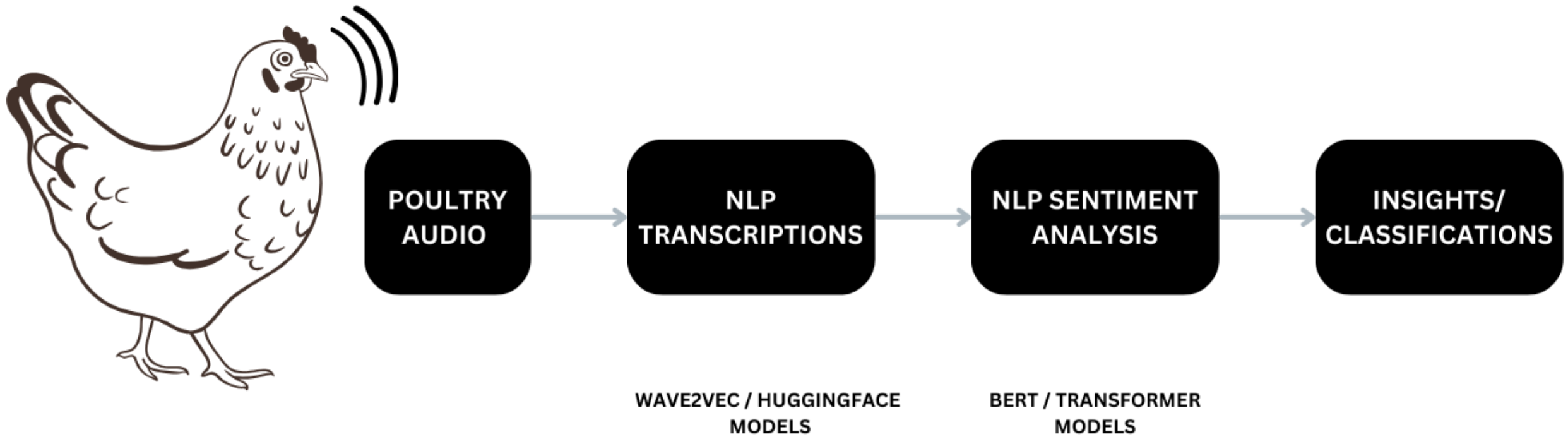

Figure 1. Overview of the Proposed Methodology for Analyzing Poultry Vocalizations. The pipeline employs Wave2Vec and Huggingface models to transcribe audio into text, followed by sentiment analysis with BERT and transformer architectures. This process yields insights into poultry vocalizations, revealing stress indicators, emotional states, and behavior patterns.

2.2. Architecture Behind Wave2Vec

Wave2Vec (Schneider et al., 2019) advances self-supervised learning of speech features directly from raw waveforms, reducing reliance on costly labeled data. Wave2Vec 2.0 (Baevski et al., 2020) refines this approach by predicting masked audio segments using contextual cues, producing robust latent representations through a CNN feature extractor that quantizes raw signals. A transformer then processes these latent units, capturing nuanced, long-range patterns in the audio. Fine-tuning the pre-trained Wave2Vec 2.0 on downstream tasks, like automatic speech recognition, requires far fewer labels. Crucially, it derives representations from waveforms directly, bypassing extensive feature engineering—an immense advantage for bioacoustics research dealing with varied species sounds. This method avoids manipulating sound duration or structure, pivotal when analyzing intricate animal calls. Wave2Vec 2.0's inclusion of transformer-based bidirectional LSTMs strengthens its capacity to model time-dependent characteristics inherent in extended audio sequences. By converting continuous signals into discretized latent variables, it generates general-purpose features that effectively handle the variability and ambient noise typical in field recordings of animal vocalizations.

2.3. The BERT Transformer

BERT (Bidirectional Encoder Representations from Transformers) revolutionized NLP by capturing bidirectional contexts of words through masked language modeling (MLM) and next sentence prediction (NSP) pre-training (Devlin et al., 2019). Unlike unidirectional predecessors, BERT leverages both left and right contexts, learning richer relational knowledge between words (Banerjee, 2024). Fine-tuning BERT on targeted NLP tasks—ranging from classification to translation—has achieved state-of-the-art results (Hamidi et al., 2024). Applying BERT's architecture to audio analysis, especially animal vocalizations, is innovative because it can model complex sequences that mirror speech. Bidirectional pre-training allows BERT to grasp dependencies among sounds by considering their full context. Pretraining BERT on extensive unlabeled animal audio fosters an understanding of general acoustic structures, reducing the labeled data needed for tasks like detecting mating calls, distress signals, or territorial vocalizations. Through MLM extended to audio—predicting masked "sounds" rather than words—BERT learns the latent sequence structure of vocalizations, accommodating incomplete or noisy recordings. By extracting nuanced temporal and contextual patterns, BERT improves both robustness and precision in analyzing challenging, real-world animal vocal data.

## 3. Materials and Methods

### 3.1. Datasets

*3.1.1. Early Disease Detection Dataset*

The early disease detection dataset was generated at Bowen University's poultry research farm (Aworinde et al., 2023). A cohort of day-old chicks was divided into two groups: one treated for respiratory conditions, the other serving as an untreated control. Each group was housed separately under controlled conditions, with microphones strategically positioned to reduce noise interference. Audio data was recorded in 24-bit samples at 96 kHz, ensuring high-fidelity capture. Data collection spanned 65 days, with recordings taken morning, afternoon, and night. Throughout, birds had free access to feed and water, mitigating extraneous stressors and ensuring vocalizations primarily reflected health status.

*3.1.2. Stress-Induced Poultry Vocal Dataset*

Building on prior work (Neethirajan, 2024a; Neethirajan, 2024b), this dataset originates from the CARUS facility at Wageningen University, the Netherlands. Fifty-two Super Nick chickens were maintained in three cages under conditions mimicking commercial production. Stress was induced by opening an umbrella and playing dog barking sounds to elicit fear responses. This approach enabled recording of stress-induced vocalizations and subsequent pattern analysis. The dataset is publicly accessible on Zenodo, facilitating broader research and validation.

*3.1.3. Chicken Language Dataset*

The chicken language dataset aims to deepen our understanding of poultry communication. Developed to support machine learning translation of chicken calls into human-interpretable information, it leverages work by Nicholas and Elsie Collias at the University of California, Los Angeles. Over 24 distinct chicken calls and their inferred meanings were documented in natural settings and annotated by an experienced poultry farmer, providing rich contextual data.

*3.2. Audio Processing Pipeline*

All collected audio underwent preprocessing via PyDub, converting files to 16 kHz mono .wav. Each file was segmented into 32 sections for more granular analysis. Wav2Vec2.0 (Hugging Face) pretrained for speech recognition produced textual transcriptions. A BERT-based model then classified each vocalization as positive, neutral, or negative.

*3.3. Phonetic and Sentiment Analysis*

Phonetic composition was examined to identify vowel/consonant patterns. Parallel processing through a thread pool allowed simultaneous handling of multiple audio files, improving efficiency. Results—including transcriptions, sentiment labels, and phonetic counts—were compiled into a JSON structure for streamlined organization and retrieval.

*3.4. Visualization and Computational Tools*

Matplotlib generated high-resolution histograms and bar charts to illustrate sentiment distributions and phonetic trends. Torch with GPU acceleration managed computationally intensive tasks, while Librosa and NLTK aided in tokenization. GPU utilization ensured parallelization and improved both processing speed and accuracy, optimizing large-scale audio data analysis.

## 4. Results and Discussion

The analysis of poultry vocalizations using NLP and transformer models has uncovered profound insights into how vocal behavior correlates with stress levels and health conditions in poultry. The vocal characteristics of poultry, including pitch, frequency, sentiment, and phonetic composition, exhibited distinct patterns depending on their physiological and emotional states. Specifically, our analysis of pitch and frequency revealed that stressed birds produced vocalizations with consistently elevated pitch and concentrated frequency ranges compared to those from relaxed birds. The Wave2Vec 2.0 model demonstrated its effectiveness in capturing these subtleties, with stressed birds showing pitch clusters around higher frequencies (e.g., 480–500 Hz), which aligns with stress-induced physiological constraints such as muscle tension and irregular respiration. In contrast, vocalizations from unstressed birds featured more dispersed frequencies, indicating a more relaxed state with greater variability in vocal output.

Sentiment analysis using a fine-tuned BERT model further elucidated the emotional landscape of poultry vocalizations. Prestress sentiment distribution revealed a relatively balanced mix of neutral, positive, and negative emotional states, reflecting everyday interactions and mild environmental challenges. In contrast, poststress vocalizations demonstrated an increase in negative sentiment and a corresponding decrease in neutral sentiment, suggesting the lingering impact of stress on the birds' emotional state. These insights are particularly significant for animal welfare, as they point to the utility of sentiment analysis in real-time welfare assessment and stress detection.

Phonetic composition analysis provided another layer of understanding, revealing that the vocalizations of stressed birds were dominated by short, staccato-like sounds—characterized by frequent use of consonants—which are consistent with abrupt and sharp vocal patterns. In contrast, vocalizations from relaxed birds featured a higher proportion of vowels, which contributed to more melodic and fluid vocal patterns. Such differences suggest that stressed birds produce vocalizations with less tonal complexity, possibly due to physiological limitations or as a behavioral adaptation to conserve energy. The prominence of consonants in stressed vocalizations also indicates increased vocal harshness, which may be used to communicate distress or alert flock members.

These changes in vocal behavior have important physiological and behavioral implications. The observed increase in pitch and reduction in vocal variety under stress is likely tied to heightened muscle tension in the laryngeal region and increased respiratory rates, both of which are typical physiological responses to stress in birds. The narrowing of pitch frequency and reduced phonetic diversity observed in poststress vocalizations may also represent an adaptive strategy to conserve energy, ensuring that vocalizations are less metabolically demanding. Behaviorally, stressed birds might alter their vocal patterns to signal distress to other flock members, potentially promoting group cohesion and mutual support in challenging environments.

Our findings demonstrate that vocal behavior serves as a reliable indicator of stress and health conditions in poultry, providing valuable insights for advancing poultry welfare through non-invasive monitoring. By integrating these advanced NLP models into precision livestock management systems, farmers can enhance their ability to monitor the well-being of their flocks, detect stress in real-time, and implement timely interventions—ultimately improving animal welfare, productivity, and sustainability in poultry farming.

### 4.1. Pitch Analysis During Stress Phases

The prestress pitch analysis (Figure 2a) highlights a widespread distribution of pitch frequencies, predominantly centered around 500 Hz. The pitch histogram reflects a relatively broad spectrum of vocal frequencies, indicative of a calm and steady state where vocal emissions are produced freely. The absence of any distinct clustering within the frequency range further implies that vocalizations during the prestress phase are primarily casual and routine.

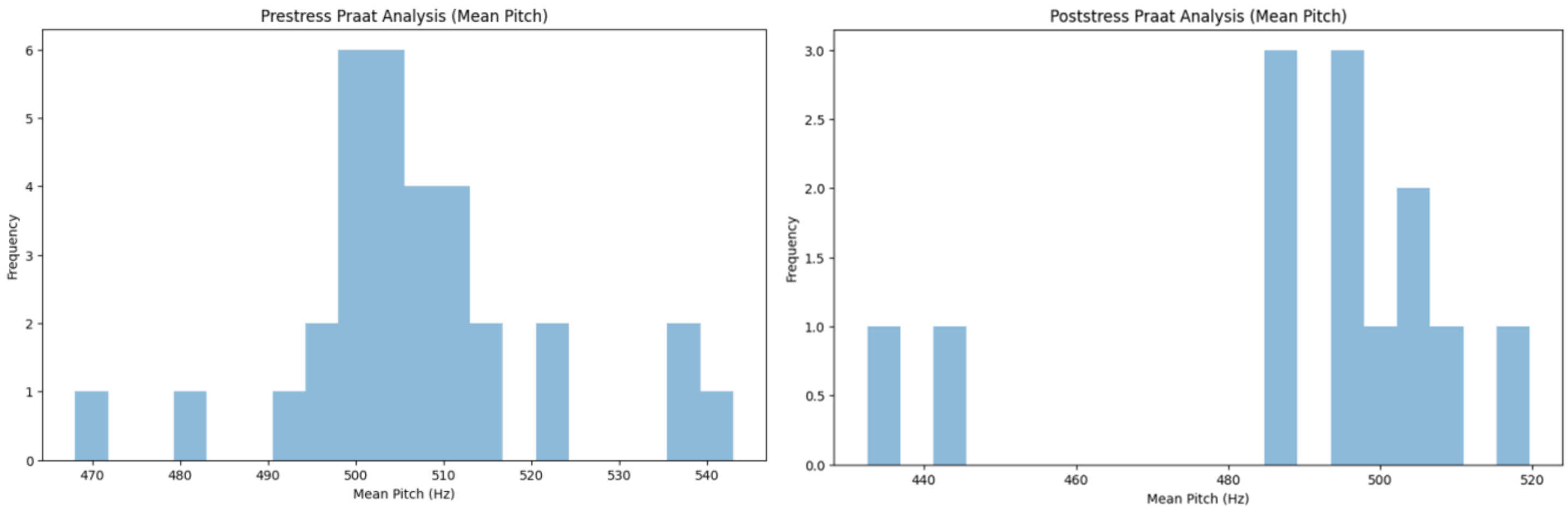

Figure 2. (a) Prestress Pitch Analysis. Distribution of pitch frequencies during the prestress phase, highlighting vocal behavior patterns under non-stressful conditions. (b) Poststress Pitch Analysis. Distribution of pitch frequencies during the poststress phase, indicating changes in vocal behavior following stress.

This suggests that the chickens are in a comfortable, non-threatening environment. Such vocal behavior likely denotes communication with peers, encompassing a range of vocal expressions for daily activities like feeding, resting, and exploring. In contrast, the poststress pitch analysis (Figure 2b) indicates a significant narrowing of the pitch frequency range, with clustering centered between 480-500 Hz. This reduction in pitch variety points toward a physiological constraint on vocalization, likely resulting from stress-induced muscle tension in the vocal apparatus. The narrowing of frequencies implies that the chickens are adjusting their vocal efforts post-stress, producing more homogenous vocalizations. Such vocal behaviors may reflect a combination of reduced airflow through tightened laryngeal muscles and a conscious attempt to conserve energy. This trend aligns with the concept of adaptive vocal suppression, wherein animals under stress reduce vocal output to minimize resource expenditure or avoid drawing attention from potential predators.

### 4.2. Physiological and Behavioral Implications of Stress-Induced Vocal Changes

The changes in pitch frequencies between prestress and poststress phases carry crucial physiological and behavioral implications. Physiologically, stress can induce significant muscle tension, particularly in the laryngeal muscles responsible for pitch control. During stress, these muscles may become tense, reducing the ability of the vocal cords to modulate pitch effectively, resulting in a narrower frequency range. Additionally, stress alters respiratory patterns, leading to shallow and irregular breathing that affects the airflow needed for a diverse pitch output. This constraint reduces the richness and complexity of vocalizations, resulting in a monotonic vocal profile characteristic of stress.

Behaviorally, such vocal adjustments may serve adaptive purposes. In stressful conditions, animals often modify their communication strategies to signal distress to others in the group. Clustering around mid-range frequencies in the poststress phase suggests that chickens may be consciously attempting to produce more distinct or easily perceivable sounds that can be detected by flock members. This serves to stimulate group cohesion or elicit protective responses from others in the flock. The reduction in vocal effort may also serve as an energy-saving mechanism, as producing varied vocalizations requires substantial metabolic effort, which could be redirected to other essential survival functions under stressful conditions.

### 4.3. Sentiment Analysis in Prestress and Poststress Phases

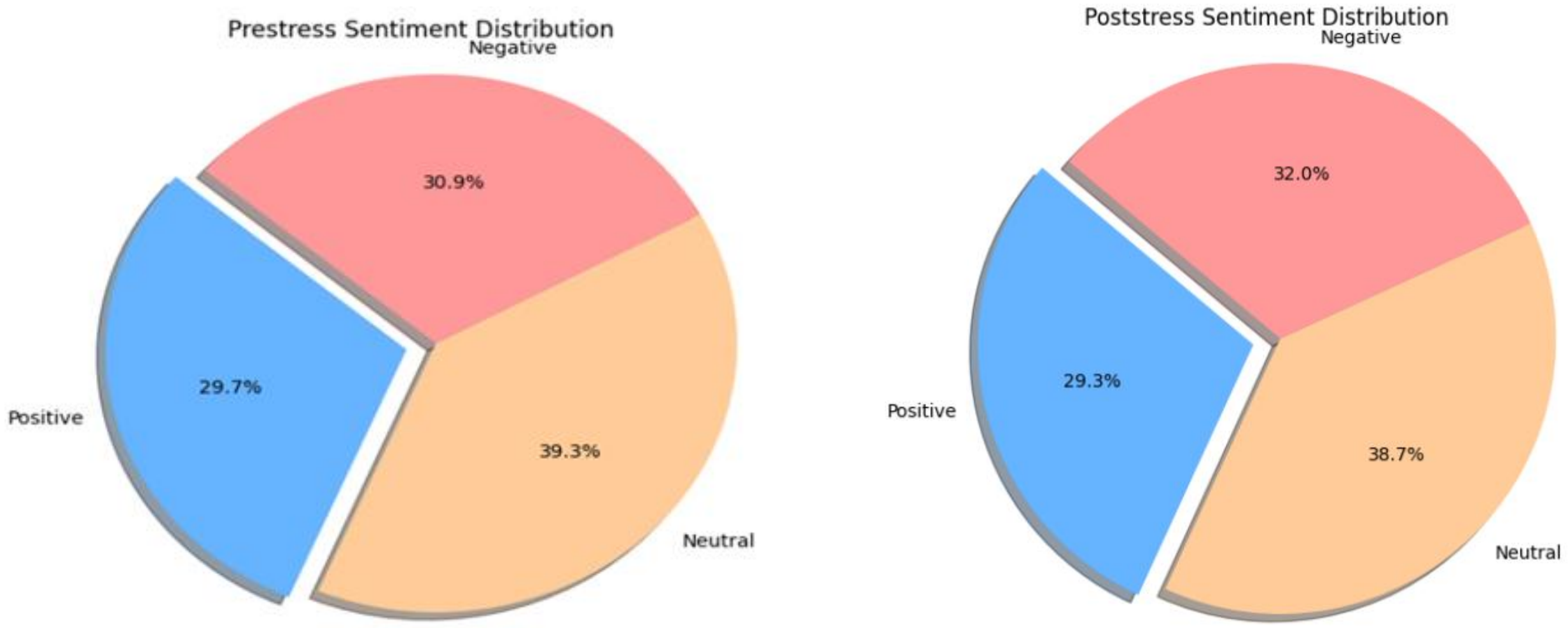

**(a)** **b)**

Figure 3. (a) Prestress Sentiment Distribution. Sentiment analysis of poultry vocalizations during the prestress phase, showing the proportions of neutral, positive, and negative sentiments. (b) Poststress Sentiment Distribution. Sentiment analysis of poultry vocalizations during the poststress phase, illustrating emotional shifts after experiencing stress.

Figure 3a illustrates the sentiment distribution during the prestress phase. Neutral sentiments constitute the largest proportion, at 39.3%, suggesting that a considerable number of vocalizations are emotionally neutral. These may include typical, non-stressful vocal expressions, such as those related to exploration or socialization. The nearly equal distribution of positive (30%) and negative (30.7%) sentiments reflects a balanced emotional state, possibly characterized by routine interactions that oscillate between minor challenges and positive stimuli. The presence of negative sentiment implies that some environmental stressors might have been present, though not at a level that would induce significant stress. In the poststress phase, sentiment distribution (Figure 3b) shows a slight shift compared to prestress. Neutral sentiments decrease slightly to 38.7%, indicating a change in the emotional tone post-stress. Negative sentiment increases to 32%, which may reflect residual stress effects or vocalizations related to stress recovery. The positive sentiment remains largely unchanged at 29.3%, implying some level of resilience among the chickens, as they attempt to revert to a normal emotional state. This indicates that, despite experiencing stress, the chickens are able to maintain a relatively balanced emotional output, possibly aided by supportive environmental factors post-stress.

## 4.4. Word Frequency Analysis in Stress Dataset

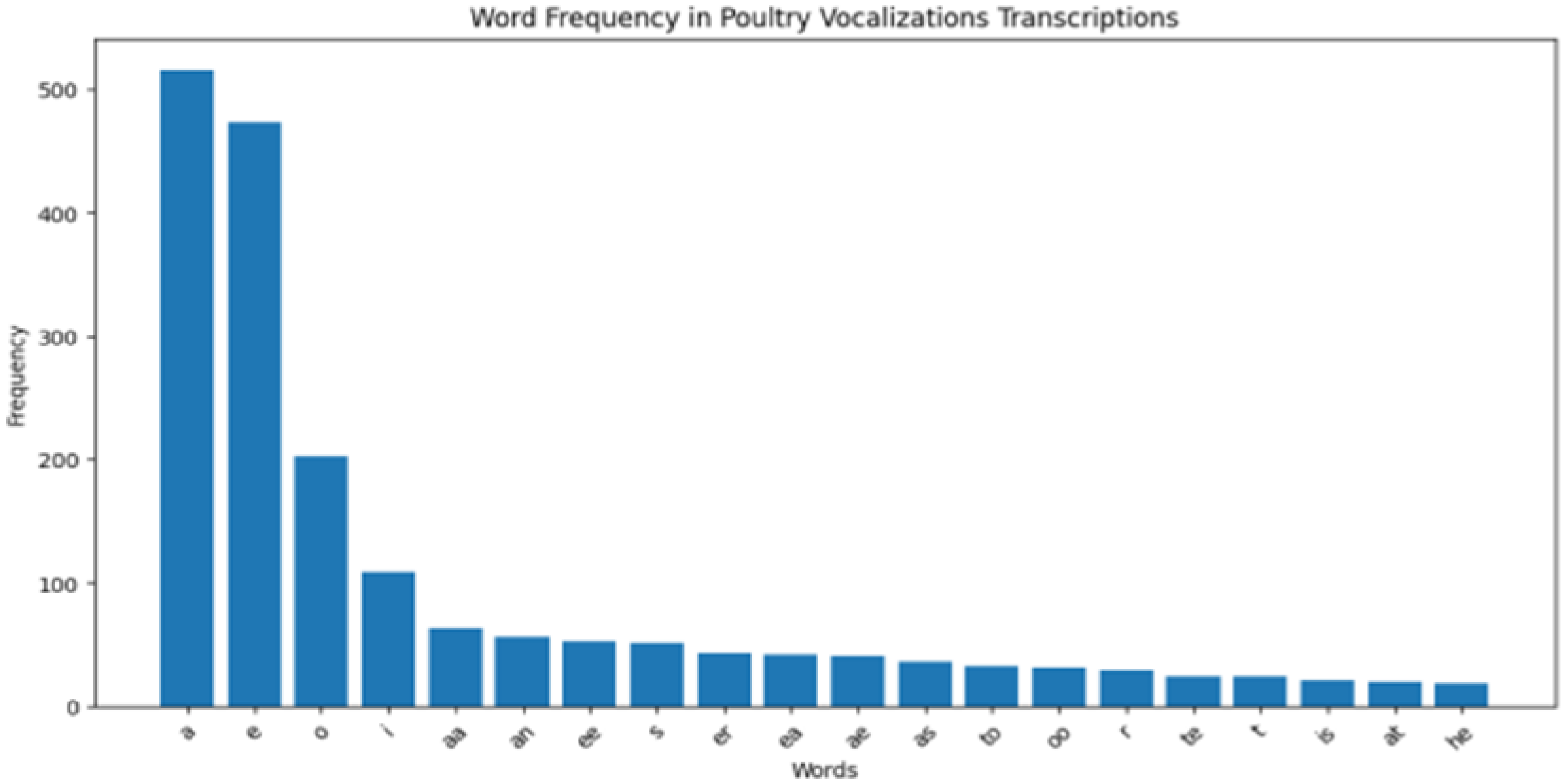

Figure 4. Word Frequency Distribution. Frequency distribution of words in poultry vocalizations, highlighting commonly occurring terms and revealing the variability in vocal repertoire.

The word frequency distribution, as depicted in Figure 5, shows a skewed pattern where a few words dominate the vocalizations, while many others are used much less frequently. The most frequent words, each occurring over 500 times, reflect simple and common vocal expressions likely used for basic communication or signaling. The sharp decline in frequency for other words suggests that the chickens employ a wide vocabulary, but most words are situational or context specific. The presence of a long tail of infrequent words highlights the complexity of chicken vocalizations and suggests that their vocal repertoire is adapted to address a wide range of scenarios, even if certain sounds are only used occasionally.

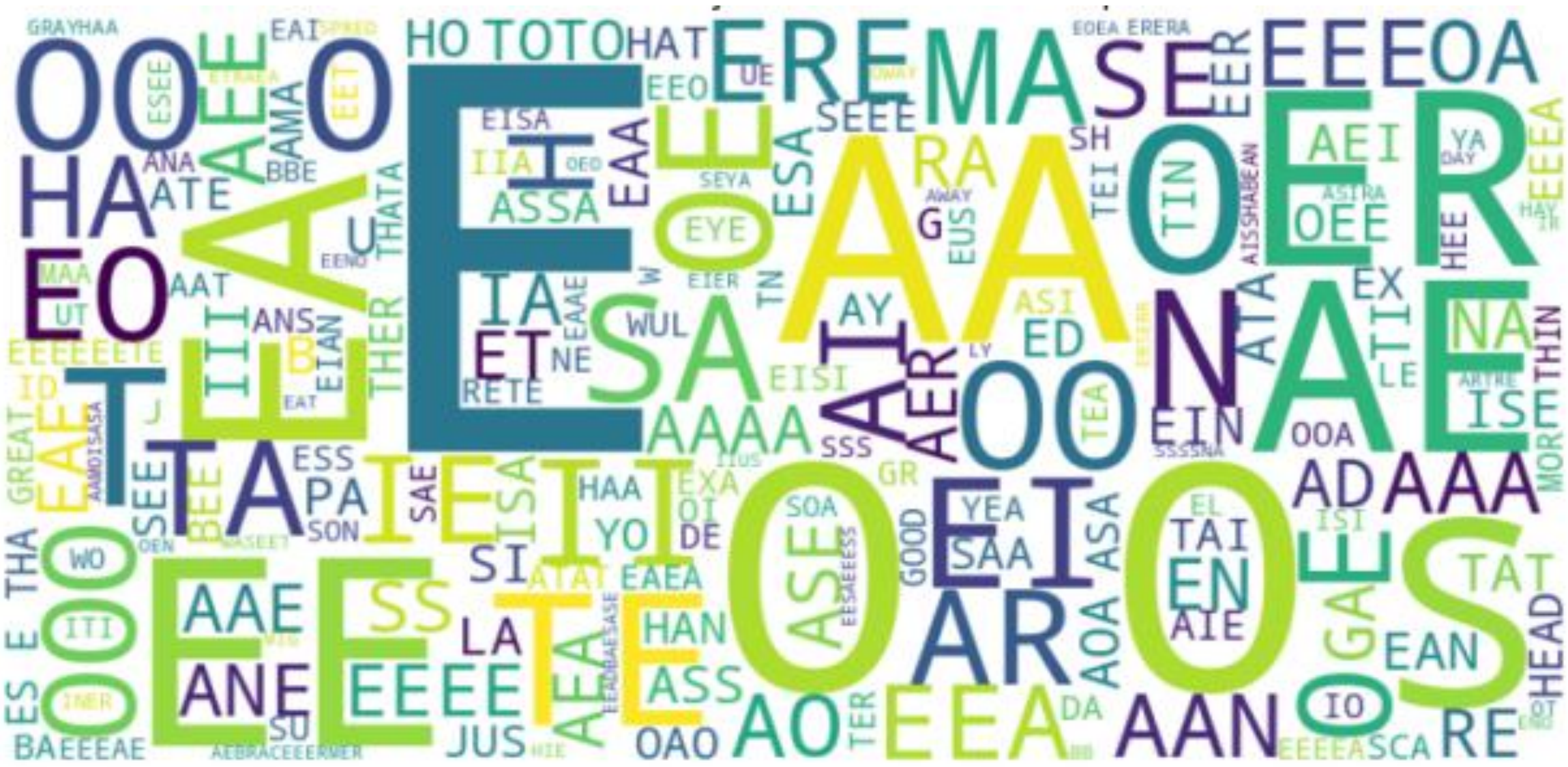

Figure 5. Word Cloud of Poultry Vocalizations. Word cloud representation of frequently used vocal elements, emphasizing dominant vowel sounds and highlighting the repetitive nature of poultry vocalizations.

The word cloud (Figure 5) further illustrates the dominance of certain vowel-rich vocalizations, such as "AAA", "EEE", and "OO", across different emotional states. These frequent vocal elements are likely the core components of poultry communication, representing simple calls that are universally understood within the flock. The presence of smaller, less frequent vocalizations in the word cloud indicates more specific or nuanced communications, which are possibly tied to particular emotional states or environmental stimuli. The redundancy of the most common vocal elements may point to their importance in ensuring effective and consistent communication within the flock, especially in the context of daily routines or social interactions.

### 4.5. Phonetic Composition Analysis

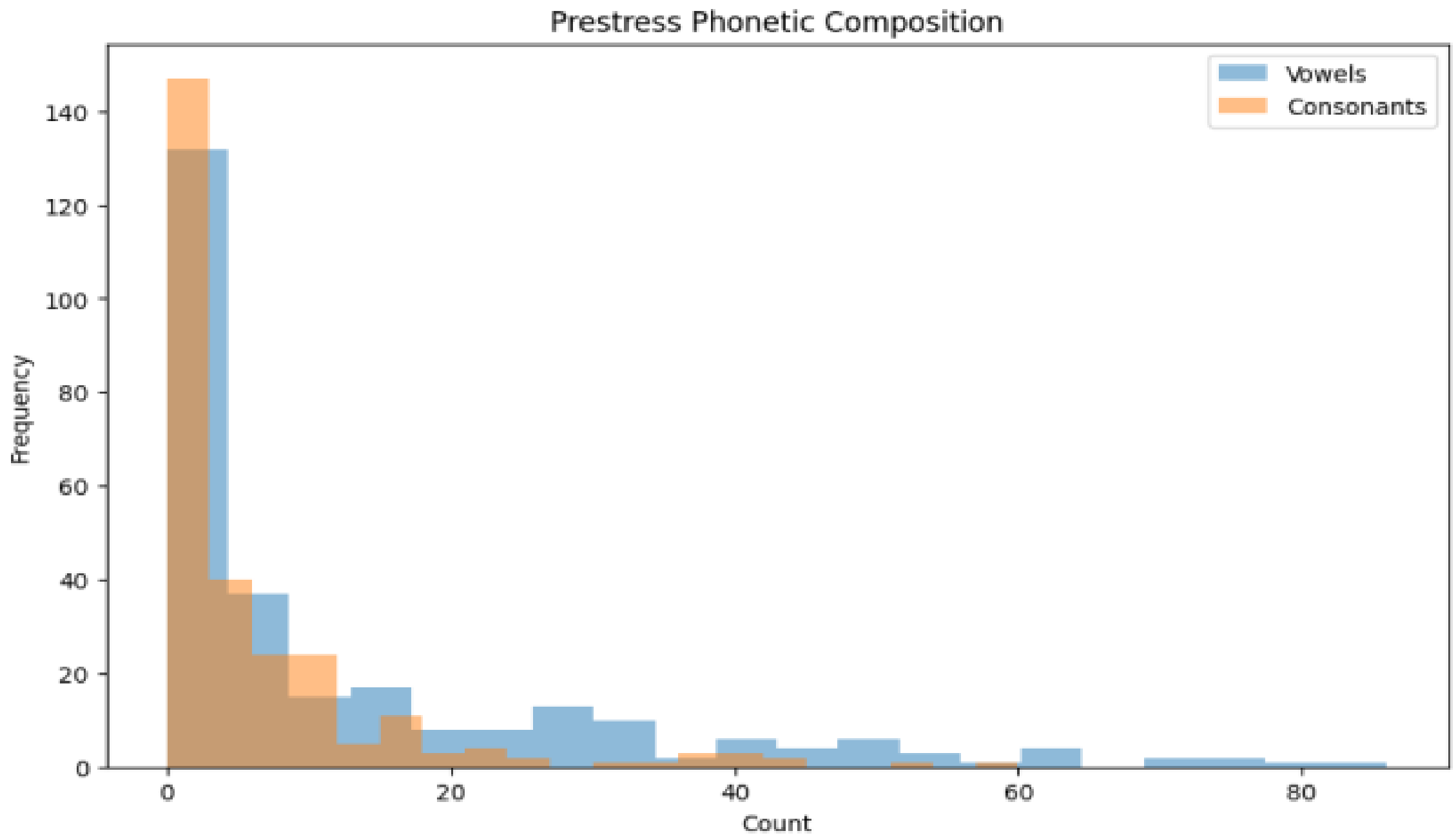

Figure 6. Prestress Phonetic Composition. Analysis of the phonetic composition of poultry vocalizations during the prestress phase, showcasing the use of vowels and consonants in vocal patterns.

Figure 6 presents the phonetic composition of vocalizations during the prestress phase, showing a bimodal distribution with clusters for both vowels and consonants. The clustering at lower counts for vowels and consonants suggests the frequent use of shorter, concise vocal expressions during the prestress phase. The overlap between vowel and consonant distributions indicates that while many vocalizations are brief, there are selective instances of more complex expressions that may involve vowel-rich vocalizations. Such vowel dominance may suggest an element of tone modulation, used to convey varying degrees of intensity or intent.

### 4.6. Phonetic Makeup During Poststress Phase

In the poststress phase, phonetic analysis reveals a shift toward vowel dominance, particularly at higher frequency levels (20 and above). This indicates a shift in vocal strategy, possibly reflecting an emotional adjustment after the stress event. The increased presence of vowels, often associated with musical or melodic qualities, may indicate an attempt to produce more soothing and less jarring vocalizations, which could contribute to emotional regulation or group cohesion post-stress. The reduction in consonants during this phase suggests a move away from harsh or staccato sounds, aligning with the idea that poststress vocalizations are adapted to reduce vocal harshness and foster a calming effect.

### 4.7. Prestress vs. Poststress Phonetic Composition

The comparative analysis of phonetic composition between prestress and poststress phases reveals several notable trends. The increased reliance on vowels poststress suggests an emotional transition towards relaxation, while the reduction in consonants may indicate a decrease in vocal harshness and complexity. Consonants, typically associated with sharper sounds, decrease in frequency, which may imply a move towards less urgent and more melodic communication. The presence of scattered outliers in vowel and consonant counts during the prestress phase points to a diverse vocal repertoire that may be associated with higher emotional arousal or variability in vocal expression.

### 4.8. Pitch Analysis and Physiological Implications

#### *4.8.1 Prestress and Poststress Pitch Characteristics*

Pitch analysis provides valuable insights into how chickens modulate their vocalizations in response to stress. During the prestress phase, vocalizations display a wide frequency range, suggesting that chickens are in a relaxed state with the freedom to express themselves across a broad tonal spectrum. In contrast, the narrowing of pitch frequency in the poststress phase indicates a physiological or behavioral adaptation to stress, characterized by reduced vocal complexity and energy conservation. This adaptation may be attributed to increased tension in the vocal muscles, limiting pitch modulation and airflow.

#### Behavioral and Social Implications

From a behavioral standpoint, changes in pitch frequency between the prestress and poststress phases suggest a conscious adjustment in vocal strategy. Chickens may reduce their vocal variety as a response to stress, possibly as a means of signaling distress to their flock members or conserving energy. Such changes in vocal behavior may also have social implications, fostering group cohesion or eliciting protective responses from the flock. The clustering of pitch frequencies in the poststress phase may indicate an attempt to produce louder or more distinct sounds, which can be easily perceived by others, thereby enhancing communication within the group during stressful situations.

**Sentiment Analysis During Healthy vs. Unhealthy Conditions**

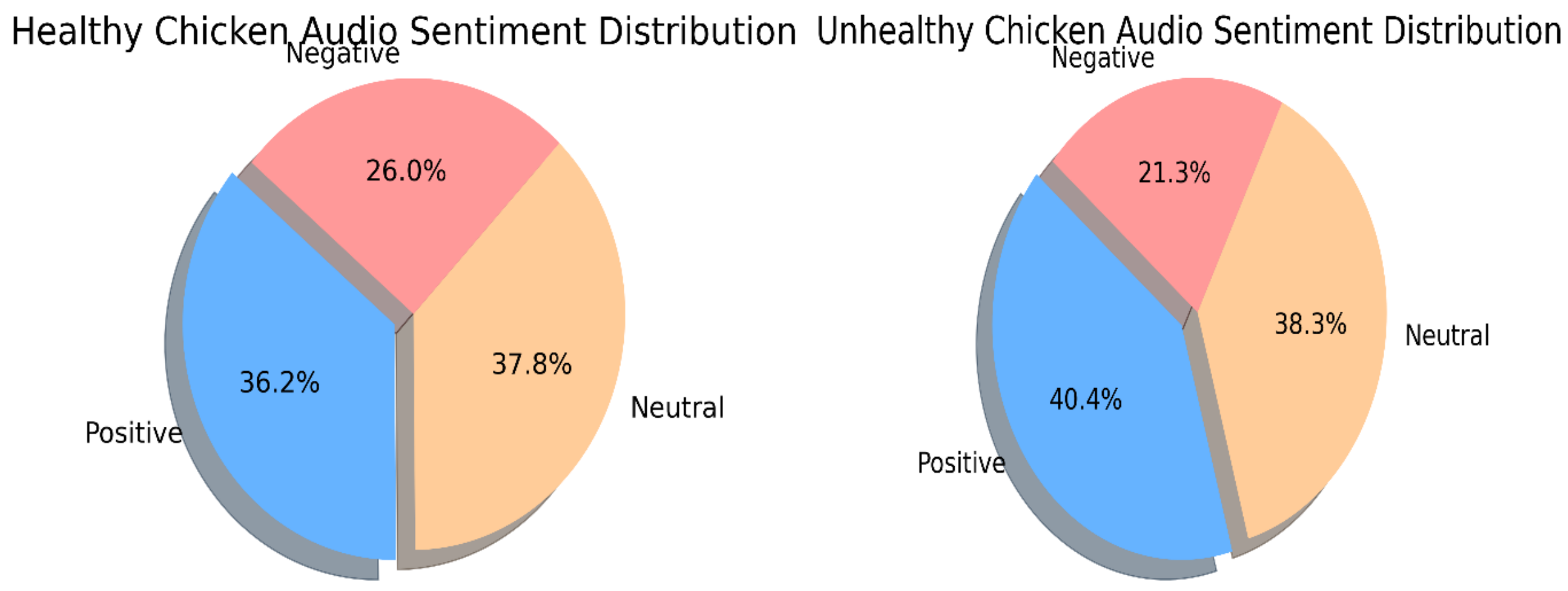

(a) (b)

Figure 7. (a) Sentiment analysis of poultry vocalizations during the Healthy condition, showing the proportions of neutral, positive, and negative sentiments. (b) Sentiment analysis of poultry vocalizations during unhealthy conditions, illustrating emotional shifts after experiencing diseased condition.

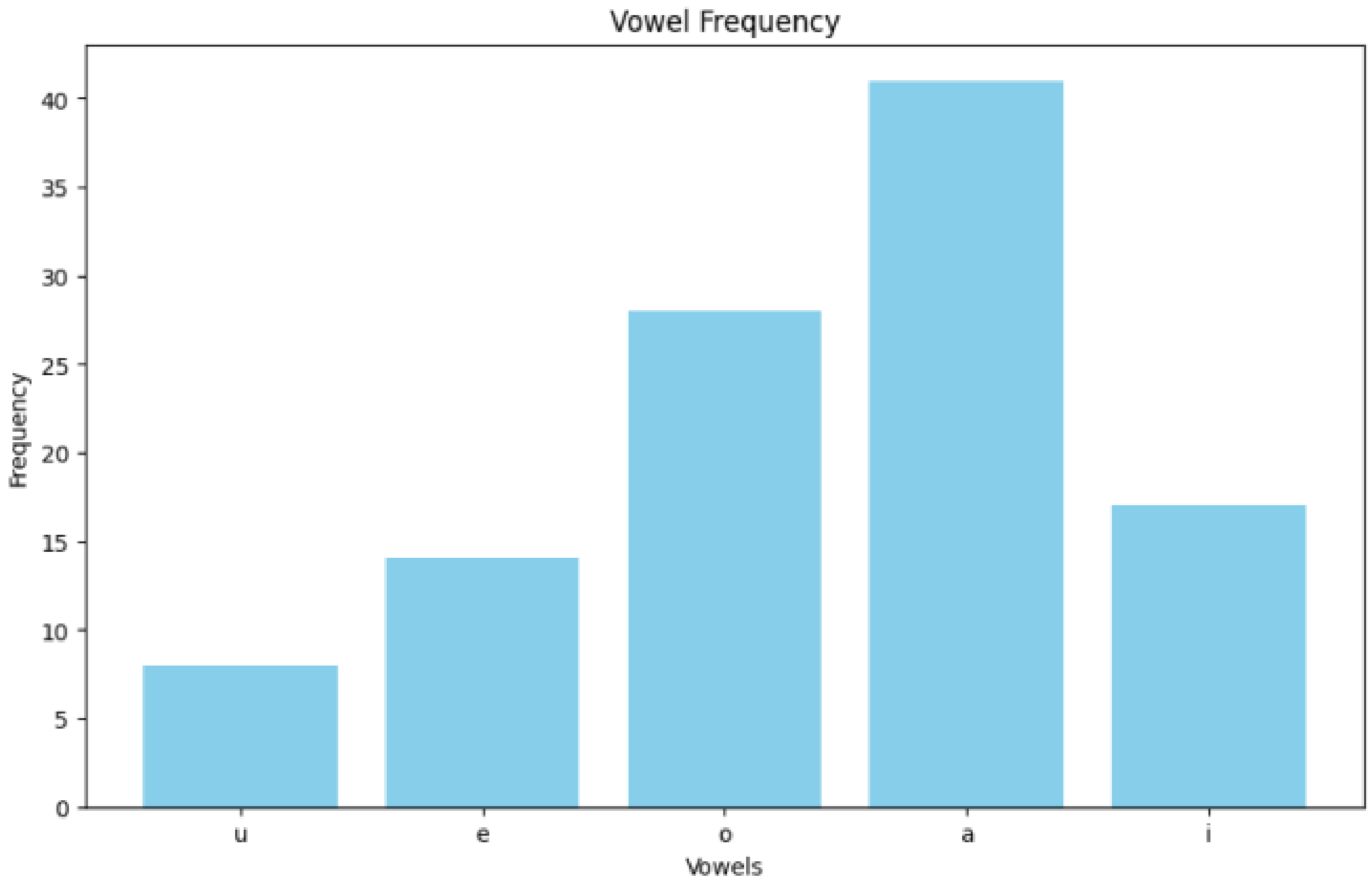

Figure 8. Vowel Frequency Comparison (Healthy vs. Unhealthy). Comparison of vowel frequency between healthy and unhealthy poultry, showing the differences in vocalization patterns based on health status.

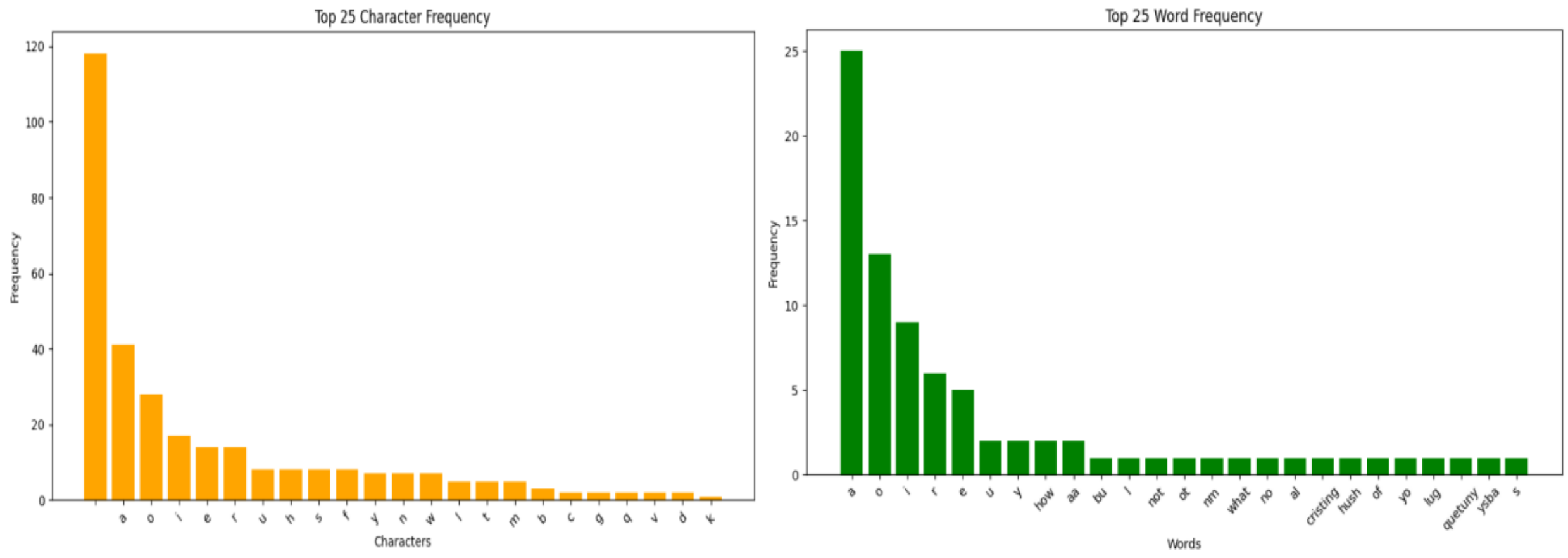

(a) (b)

Figure 9. (a) Character Frequency Distribution in Poultry Vocalizations. This figure illustrates the top 25 most frequently occurring characters in poultry vocalization transcriptions. The dominant character 'a' highlights its prevalence across vocalizations, indicating a recurring phonetic element that may be crucial for basic communication patterns in poultry. (b). Word Frequency Distribution in Poultry Vocalizations. This figure presents the top 25 words used in the analyzed poultry vocalizations, revealing a significant presence of simple vowel-rich words like "a" and "o." The distribution suggests that poultry vocal communication is largely composed of basic, repetitive sounds, reflecting both simplicity and functionality in their vocal repertoire

The comparison of vowel frequency in healthy and unhealthy chickens (Figure 8) reveals that vowels "a", "o", and "e" dominate in both conditions, indicating their fundamental role in poultry vocalization. The increased variation in vowel frequency in unhealthy chickens suggests greater vocal effort, possibly indicating discomfort or a heightened attempt to communicate distress. The dominance of specific vowels, even under different health conditions, highlights the importance of these sounds in the basic structure of poultry communication.

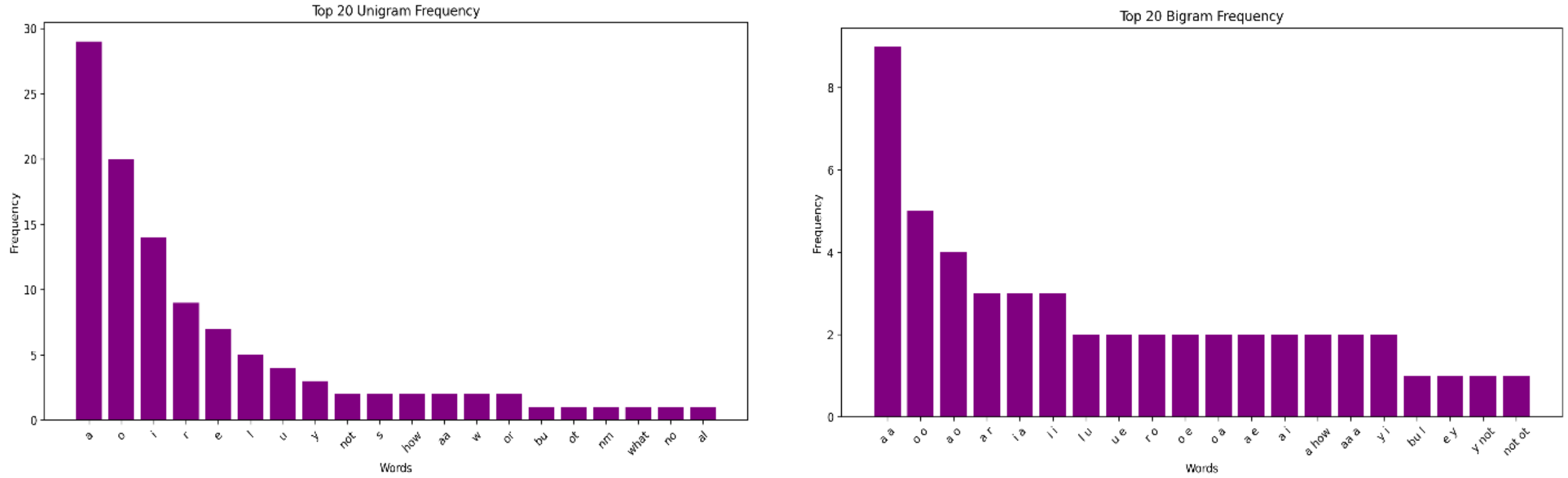

(a) (b)

Figure 10. (a) Unigram and (b) Bigram Frequency Distribution (Healthy vs. Unhealthy). Unigram and bigram frequency distribution comparing healthy and unhealthy poultry vocalizations, providing insight into the complexity of vocal expressions.

Unigram analysis (Figure 10a) shows that "a" is the most common sound across both healthy and unhealthy chickens, followed by "o", "i", and "e". The increased variation in unhealthy conditions suggests a more diverse vocal effort, possibly indicating an increased need to express discomfort. Bigram analysis (Figure 10b) provides further insights, with repeated phonemes like "aa" being prominent in stressful conditions, which may indicate an intensified attempt to communicate urgency or distress. The variety of bigrams in unhealthy conditions suggests that stress or illness prompts chickens to experiment with different vocal patterns, increasing the diversity of vocal expressions.

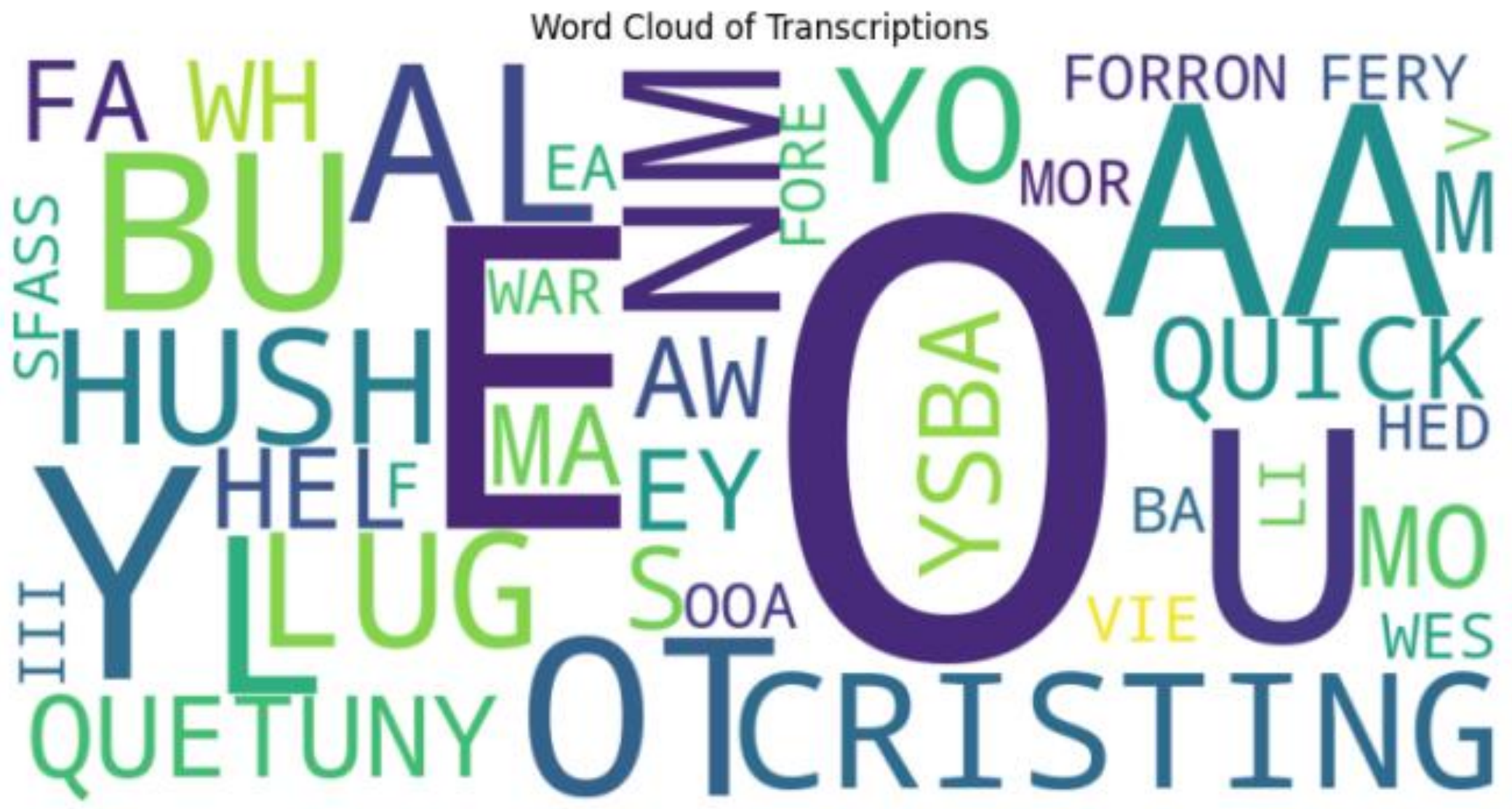

Figure 11. Word Cloud Analysis (Healthy vs. Unhealthy). Word cloud depicting vocal elements of healthy vs. unhealthy poultry, emphasizing differences in vocal patterns linked to health status.

The word cloud (Figure 11) presents a visual representation of the vocal patterns in healthy versus unhealthy chickens. Simple vowel-rich words such as "O", "AA", and "E" are predominant in healthy chickens, indicating a reliance on basic vocal elements for everyday communication. In contrast, sounds like "LUG", "BU", and "HUSH" are more frequently used in unhealthy chickens, suggesting heightened use of context-specific vocalizations related to stress or discomfort. This difference highlights the adaptability of poultry vocalization in response to changing health conditions.

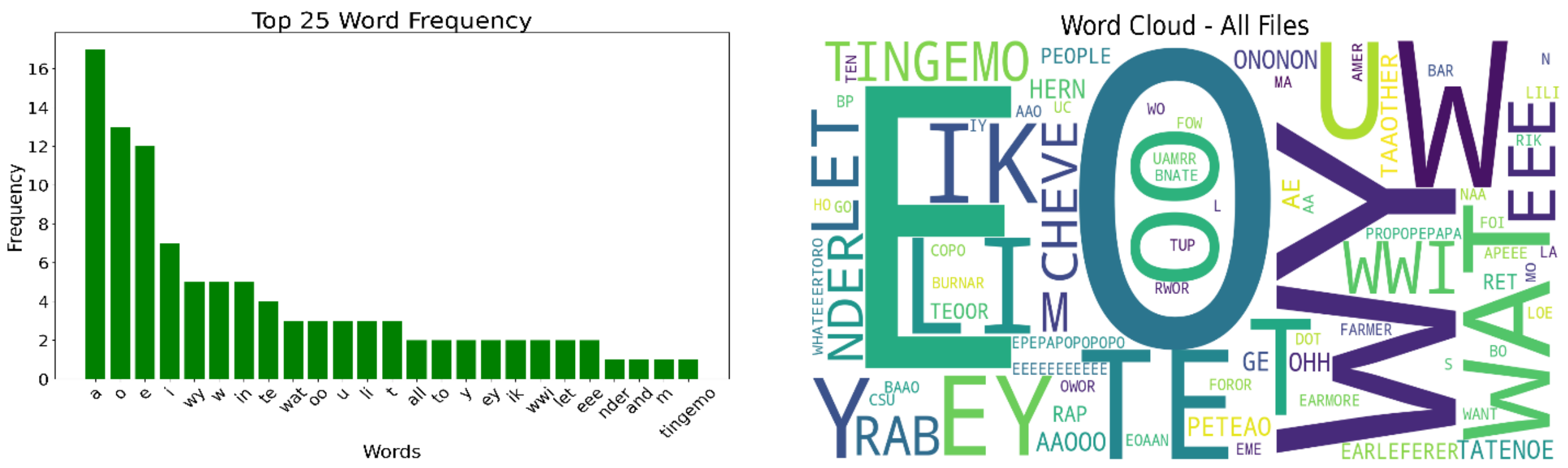

Figure 12. (a) Top word Frequency in the Chicken Language Dataset. We can see asimilarity with the other datasets  (b) Word cloud Diagram of the same dataset. We are able to observe new words in this cloud set.

On analysis of the third dataset – the Chicken language dataset, we find somewhat similar results as well. The Chicken Language Dataset presents an intriguing perspective on how chickens interact with each other. This resource was compiled by Nicholas and Elsie Collias by observing the chickens in a farm, through which, they were able to capture the diversity in chicken vocalization and language. In various illustrations included in this work, its analysis opens important patterns for vowel usage and frequency of different words and characters as well as overall complexity of linguistics involved.

The vowel frequency chart shows the prominence of certain vowels in the utterances made by the chickens. The instance of vowel “e” is the highest followed by “o” and “a,” thus making them the notable triad in chicken communication. Vowels such as “i” and “u” however, are there though relatively in a lower instance and may have supporting roles or be used for less frequent sounds. The imbalance in instance dominance of the two groups in the chicken language could be due to some physiological influencing or limiting factors as well as the relevance of the vowels in the pronouncement of various emotional and environmental signals.

The word frequency distribution offers more detailed analysis revolving around the calls made by the chickens. Vowel words or calls “a”, “o”, “e” have once again been highlighted among the primary vocabulary of chickens which is evident in the vocalization chart. Certain expressions in chickens’ interactions with each other such as “wy” or “in” might imply a sequence of sequential vocalization such as providing an alarm about food or danger. The extensive variety of words suggests that chickens might have a blocked form of language which is systematized to suit various situations.
The character frequency chart also depicts how language is structured for the calls made by chickens. The letter e is reported to be the most common character, being followed by a good number of occurrences of o, a, and t. These characters are forming the calls of the chicken conforming to vowels being in the majority in the dataset for this study. It is inferred that c, and d among others are less predominant or less essential characters, therefore they may correspond to limitation in explosives or environmental effects accompanying the production of sound.
The word cloud provides extensive information over the dataset as well as the most appropriate words and their behaviours lived up to those frequencies. The terms “T”, “INGEMO” and “EIK”, and “WWI” are some of the notable terms demonstrating a pattern of engagement displays with a likely meaning in communication among chickens.

The high frequency of vowels used, the definite occurrence of the same words and characters in the middle of others as well as the degree of variability in the word cloud suggest that the system is indeed complex, robust and situational. Such results emphasize the possibility of further development of audio analysis of the hen's calls, including training of artificial neural networks for automatic translation of hen's calls. By correlating specific calls with specific emotions or external influences, they can help improve animal welfare monitoring systems and deepen knowledge of communication systems in other species.

The sociological perspective provided by the study of the chicken language dataset is striking: there are intricate patterns of social and communicative behavior among chickens. The varied sounds are used to fulfill the purpose of staying in the group and for some, the sound allows the chicken to find the other chickens who are somewhere in the area. Such sounds allow the chicken to feel safe within the group and help in keeping order. There are even certain sounds that do or direct which members of the group should follow the other ones just making the group even stronger and less open to threats.

There also seems to be some level of role differentiation among the in-group members as demonstrated by the vocal patterns. The calls refer to behaviors differing among roosters, chickens and even chicks for they relate communication with their role in an appropriate manner. Similarly, sounds produced when chicks are being guided or mates are being called out imply that there is seniority and role division in the group. Such differentiation enhances efficiency in the operations of the social system and showcases the advanced social understanding in relations onto roles.
Another key aspect includes regard and sensitivity to emotions and situation. There are other sounds including the ones a chicken makes when they are hungry, thirsty, uncomfortable, or are feeling heat which promotes a need to bring the sound with the state they wish to project.

## Pitch Analysis - Physiological and Behavioral Adaptations

### Pitch Analysis During Healthy vs. Unhealthy Phases

Pitch analysis provides critical insights into how stress and health conditions influence vocal behavior. In healthy chickens, the pitch frequency distribution is broad, centered around 500 Hz, with vocal emissions dispersed across a wide spectrum. This wide distribution suggests that chickens are able to vocalize freely and express a wide range of emotions. In contrast, in unhealthy chickens, the pitch frequency distribution is significantly narrower, reflecting physiological or behavioral adaptations to discomfort or illness. This reduction in pitch variety may indicate increased muscle tension, reduced airflow, and adaptive vocal suppression.

## Training and Model Performance Analysis

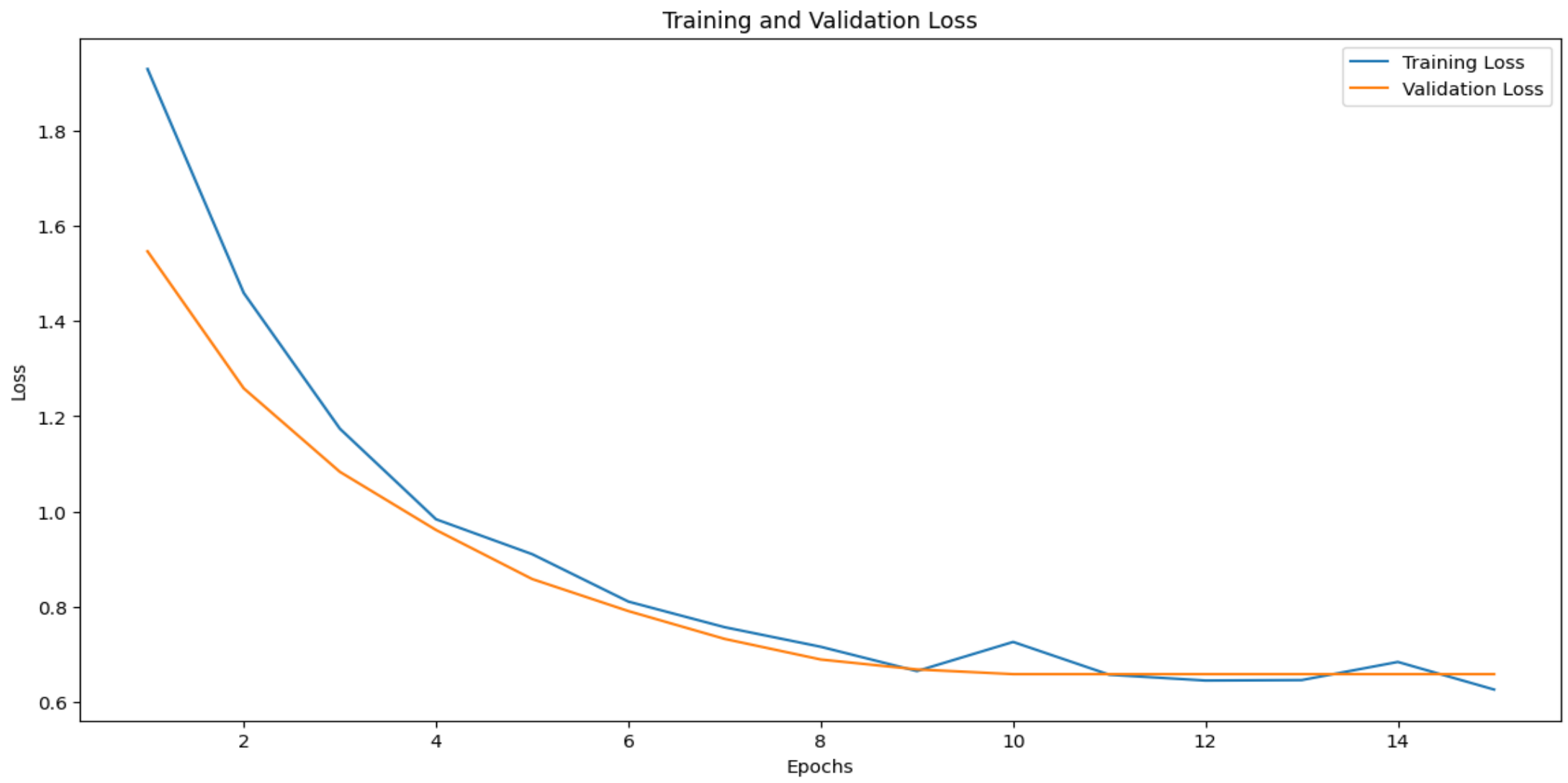

Figure 13. Training vs. Validation Loss Over 15 Epochs. Plot of training versus validation loss over 15 epochs, illustrating the model's convergence and generalization during training.

Figure 13 illustrates the training and validation loss over 15 epochs for the BERT-based model. The consistent decrease in both training and validation loss demonstrates the model's effective learning capabilities. The convergence of these metrics suggests that the model generalizes well without overfitting, largely due to the use of dropout regularization and a learning rate scheduler. The steady decline in loss metrics across epochs indicates that the model effectively learns meaningful representations of poultry vocalizations, capturing the nuanced differences in emotional states.

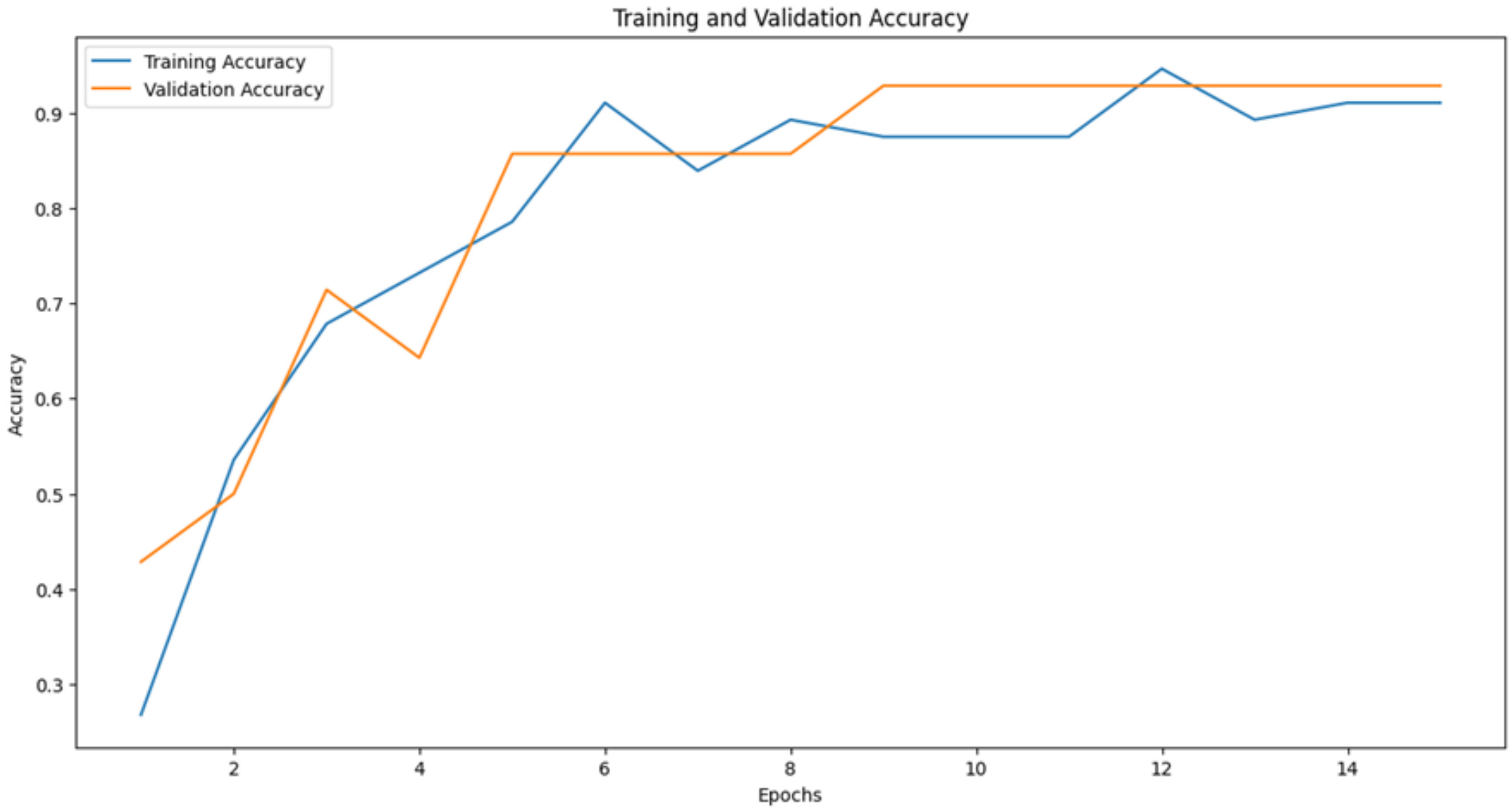

Figure 14. Training vs. Validation Accuracy Over 15 Epochs. Plot of training versus validation accuracy over 15 epochs, showing the model's performance in classifying poultry vocalizations.

Figure 14 presents the accuracy metrics during training and validation. Both metrics improve steadily over epochs, stabilizing at around 93% for validation accuracy. This stability, combined with consistent declines in loss metrics, highlights the robustness of the transformer-based architecture in learning from complex vocal data. The alignment of training and validation metrics suggests that the model captures the essence of stress-induced vocal variations, reinforcing its applicability in real-world scenarios where vocal behavior monitoring is crucial.

**Leveraging Research Insights for Poultry Farming Management**

The application of NLP and transformer models to poultry vocalizations presents transformative potential for improving welfare monitoring and management practices on a large scale. The insights gained from pitch, sentiment, and phonetic analyses provide a foundation for real-time welfare monitoring that can alert farmers to stress-induced vocalizations, enabling timely intervention. The detection of changes in vocal patterns may serve as early indicators of stress or illness, supporting proactive measures to improve flock health.

**Environmental and Resource Management**

Establishing correlations between vocalization patterns and environmental factors such as temperature, lighting, or feed can enhance farm management strategies. For instance, detecting specific calls associated with elevated temperatures can prompt temperature regulation, reducing heat stress. Understanding how vocalizations vary with lighting conditions may allow farmers to optimize lighting regimes to improve poultry welfare and productivity. Similarly, analysis of vocal changes related to feed introduction could provide insights into palatability or nutrient absorption, enabling refinements in nutritional programs.

**Social and Behavioral Management**

The vocalization analysis also provides insights into the social dynamics within the flock. Monitoring changes in vocal behavior during flock integration or assessing vocal patterns related to aggression can help farmers manage social interactions effectively, reducing stress and injury. Identifying vocal cues associated with mating behaviors could enhance breeding programs, increasing reproductive success. Integrating vocalization data with precision livestock technologies, such as weight and movement sensors, could facilitate accurate predictive models for growth, health, and welfare outcomes, ultimately improving resource allocation and sustainability in poultry farming.

**Towards Precision Livestock Farming**

Incorporating vocalization analysis into broader precision livestock farming (PLF) frameworks offers opportunities for significant advancements in animal welfare and farm efficiency. The integration of vocalization data with sensor-based monitoring systems could enable more comprehensive assessments of animal health and well-being. For example, combining vocal data with accelerometer (or video-based activity monitoring) data could offer a fuller picture of both the physical and emotional states of poultry, leading to enhanced decision-making capabilities. Leveraging insights from vocalizations, such as recognizing early stress indicators, allows for optimized use of resources, reducing waste and promoting sustainable farming practices. The findings, from pitch and phonetic analyses to sentiment and word frequency evaluations, reveal significant insights into how chickens communicate in response to stress and health challenges.

## Conclusions

This study demonstrates the profound potential of NLP and transformer-based models—such as Wave2Vec 2.0 and BERT—in deciphering the "language" of poultry, offering a powerful new lens through which to understand avian health, stress, and emotional states. By translating subtle acoustic patterns into meaningful insights, our work reframes poultry vocalizations from background farm noise into a rich data stream reflecting welfare conditions, environmental challenges, and complex social dynamics. This shift is more than a technical advancement; it embodies a forward-thinking approach to animal husbandry that respects and responds to the cognitive and emotional realities of poultry, paving the way toward more humane, informed, and ecologically mindful farming practices.

Key takeaways are clear: Vocalization-based analysis provides a non-invasive, scalable means of anticipating stressors, detecting illnesses, and adjusting management strategies before issues escalate. Moving from reactive to proactive care, farmers can enhance animal well-being, improve productivity, and foster sustainable agri-ecosystems. As the world calls for higher welfare standards and ecological responsibility, these AI-driven methods align ethical considerations with operational efficiency.

However, to fully realize this potential, several challenges must be addressed. Future efforts should aim to improve noise-robustness, ensuring reliable data extraction in the cacophony of real farm environments. Validation across diverse breeds, farm types, and geographical regions will ensure model applicability on a global scale. Beyond chickens, adapting these techniques to other avian species—or even broader wildlife acoustic monitoring—could enrich biodiversity assessments and ecological informatics. Integrating these acoustic insights with environmental sensor networks and multimodal data (visual cues, physiological indicators, and habitat parameters) will produce a holistic view of animal health and environmental quality.

In essence, this research marks only the beginning of a transformative journey. By understanding and respecting animal "voices," we open avenues for data-driven, compassionate stewardship. As we refine these models, embrace complexity, and integrate new data streams, the promise of intelligent, ethical, and sustainable agriculture moves ever closer to reality.

**Author Contributions**

Conceptualization, S.N.; methodology, S.N.; software, V.M.; validation, V.M.; formal analysis, V.M.; investigation, S.N. and V.M.; resources, S.N.; writing—original draft preparation, V.M.; writing—review and editing, S.N.; visualization, V.M.; supervision, S.N.; project administration, S.N.; funding acquisition, S.N. All authors have read and agreed to the published version of the manuscript.

**Funding**

This work is kindly sponsored by the Natural Sciences and Engineering Research Council of Canada (RGPIN 2024-04450), and the Department of NB Agriculture (NB2425-0025).

**Data Availability Statement**

The data are available from the corresponding author upon reasonable request.

**Conflicts of Interest**

The authors declare no conflicts of interest.

**References**

Amirivojdan, A., Nasiri, A., Zhou, S., Zhao, Y., & Gan, H. (2024). "ChickenSense: A Low-Cost Deep Learning-Based Solution for Poultry Feed Consumption Monitoring Using Sound Technology," AgriEngineering, 6(3), 2115–2129. DOI: 10.3390/agriengineering6030124

Aworinde, H., Adebayo, S., Akinwunmi, A., Alabi, O., Ayandiji, A., Oke, O., Oyebamiji, A., Adeyemo, A., Sakpere, A., & Echetama, K. (2023). "Poultry Vocalization Signal Dataset for Early Disease Detection," Mendeley Data, V1. DOI: 10.17632/zp4nf2dxbh.1

Baevski, A., Zhou, H., Mohamed, A., & Auli, M. (2020). "wav2vec 2.0: A Framework for Self-Supervised Learning of Speech Representations," arXiv preprint, arXiv:2006.11477. DOI: 10.48550/arXiv.2006.11477

Banerjee, D. (2024). "Knowledge Graph Question Answering with Generative Language Models," Dissertation, Universität Hamburg. DOI: 10.5281/zenodo.6808839

Berthet, M., Coye, C., Dezecache, G., & Kuhn, J. (2023). "Animal linguistics: a primer," Biological Reviews, 98(1), 81–98. DOI: 10.1111/brv.12897

Bolhuis, J. J., Beckers, G. J. L., Huybregts, M. A. C., Berwick, R. C., & Everaert, M. B. H. (2018). "Meaningful syntactic structure in songbird vocalizations," PLOS Biology. DOI: 10.1371/journal.pbio.2005157

Bravo Sanchez, F. J., Hossain, M. R., English, N. B., & Moore, S. T. (2021). "Bioacoustic classification of avian calls from raw sound waveforms with an open-source deep learning architecture," Scientific Reports, 11(1), 15733. DOI: 10.1038/s41598-021-95076-6

Cohen, Y., Nicholson, D. A., Sanchioni, A., Mallaber, E. K., Skidanova, V., & Gardner, T. J. (2022). "Automated annotation of birdsong with a neural network that segments spectrograms," eLife, 11, e63853. DOI: 10.7554/eLife.63853

Collias, N. E. (1987). "The vocal repertoire of the red junglefowl: A spectrographic classification and the code of communication," The Condor, 89(3), 510–524. DOI: 10.2307/1368641

Coutant, M., Villain, A. S., & Briefer, E. F. (2024). "A scoping review of the use of bioacoustics to assess various components of farm animal welfare," Applied Animal Behaviour Science, 275, 106286. DOI: 10.1016/j.applanim.2024.106286

Davis, S., & Mermelstein, P. (1980). "Comparison of parametric representations for monosyllabic word recognition in continuously spoken sentences," IEEE Transactions on Acoustics, Speech, and Signal Processing, 28(4), 357–366. DOI: 10.1109/TASSP.1980.1163420

Devlin, J., Chang, M.-W., Lee, K., & Toutanova, K. (2019). "BERT: Pre-training of Deep Bidirectional Transformers for Language Understanding," Proceedings of NAACL-HLT, 4171–4186. DOI: 10.18653/v1/N19-1423

Edgar, J. L., Lowe, J. C., Paul, E. S., & Nicol, C. J. (2011). "Avian maternal response to chick distress," Proceedings of the Royal Society B: Biological Sciences, 278, 3129–3134. DOI: 10.1098/rspb.2010.2701

Fontana, I., Tullo, E., Butterworth, A., & Guarino, M. (2015). "An innovative approach to predict the growth in intensive poultry farming," Computers and Electronics in Agriculture, 119, 178–183. DOI: 10.1016/j.compag.2015.10.001

Galef, B. G. Jr., & Laland, K. N. (2005). "Social Learning in Animals: Empirical Studies and Theoretical Models," BioScience, 55(6), 489–499. DOI: 10.1002/0470018860.s00721

Gong, Y., Chung, Y.-A., & Glass, J. (2021). "AST: Audio Spectrogram Transformer," arXiv preprint, arXiv:2104.01778. DOI: 10.48550/arXiv.2104.01778

Goyal, V., Yadav, A., & Mukherjee, R. (2024). "A Literature Review on the Role of Internet of Things, Computer Vision, and Sound Analysis in a Smart Poultry Farm," ACS Agricultural Science & Technology, 4(4), 368–388. DOI: 10.1021/acsagscitech.3c00467

Gupta, G., Kshirsagar, M., Zhong, M., Gholami, S., & Ferres, J. L. (2021). "Comparing recurrent convolutional neural networks for large-scale bird species classification," Scientific Reports, 11, 17085. DOI: 10.1038/s41598-021-96446-w

Güntürkün, O. (2005). "The avian 'prefrontal cortex' and cognition," Current Opinion in Neurobiology, 15(6), 686–693. DOI: 10.1016/j.conb.2005.10.003

Hamidi, H., Hosseini, M., & Hosseyni, S. S. (2025). "Design of a Framework using Bidirectional Encoder Representations from Transformers to Understanding Panic Buying Behavior During the COVID-19 Pandemic," International Journal of Engineering, 38(1), 247–261. DOI: 10.5829/ije.2025.38.01a.22

He, P., Chen, Z., Yu, H., Hayat, K., He, Y., Pan, J., & Lin, H. (2022). "Research Progress in the Early Warning of Chicken Diseases by Monitoring Clinical Symptoms," Applied Sciences, 12(11), 5601. DOI: 10.3390/app12115601

Herborn, K. A., McElligott, A. G., Mitchell, M. A., Sandilands, V., Bradshaw, B., & Asher, L. (2020). "Spectral entropy of early-life distress calls as an iceberg indicator of chicken welfare," Journal of the Royal Society Interface, 17, 20200086. DOI: 10.1098/rsif.2020.0086

Hu, S., Chu, Y., Wen, Z., Zhou, G., Sun, Y., & Chen, A. (2023). "Deep learning bird song recognition based on MFF-ScSEnet," Ecological Indicators, 154, 110844. DOI: 10.1016/j.ecolind.2023.110844

Jarvis, E. D. (2005). "Evolution of vocal learning and spoken language," Science. DOI: 10.1126/science.aax028

Karatsiolis, S., Panagi, P., Vassiliades, V., Kamilaris, A., Nicolaou, N., & Stavrakis, E. (2024). "Towards understanding animal welfare by observing collective flock behaviors via AI-powered

Analytics," Proceedings of the 19th Conference on Computer Science and Intelligence Systems (FedCSIS), ACSIS, 39, 643–648. DOI: 10.15439/2024F2064

Laleye, F. A. A., & Mousse, M. A. (2024). "Attention-based recurrent neural network for automatic behavior laying hen recognition," Multimedia Tools and Applications, 83, 62443–62458. DOI: 10.1007/s11042-024-18241-9

Lapp, S., Rhinehart, T., Freeland-Haynes, L., Khilnani, J., Syunkova, A., & Kitzes, J. (2023). "OpenSoundscape: An open-source bioacoustics analysis package for Python," Methods in Ecology and Evolution. DOI: 10.1111/2041-210X.14196

Lev-ron, T., Yitzhaky, Y., Halachmi, I., & Druyan, S. (2024). "Classifying Vocal Responses of Broilers to Environmental Stressors via Artificial Neural Network," Animal, 101378. DOI: 10.1016/j.animal.2024.101378

Lostanlen, V., et al. (2024). "BirdVoxDetect: Large-Scale Detection and Classification of Flight Calls for Bird Migration Monitoring," IEEE/ACM Transactions on Audio, Speech, and Language Processing, 32, 4134–4145. DOI: 10.1109/TASLP.2024.3444486

Lou, P., Wu, T., Yang, S., Wu, X., Chen, J., Zhu, X., et al. (2023). "Deep learning reveals rapid vegetation greening in changing climate from 1988 to 2018 on the Qinghai-Tibet Plateau," Ecological Indicators, 148, 110020. DOI: 10.1016/j.ecolind.2023.110020

Manteuffel, G., Puppe, B., & Schön, P. C. (2004). "Vocalization of farm animals as a measure of welfare," Applied Animal Behaviour Science, 88(1-2), 163–182. DOI: 10.1016/j.applanim.2004.02.012

Marino, L. (2017). "Thinking chickens: A review of cognition, emotion, and behavior in the domestic chicken," Animal Cognition. DOI: 10.1007/s10071-016-1064-4

McGrath, N., Dunlop, R., Dwyer, C., Burman, O., & Phillips, C. J. C. (2017). "Hens vary their vocal repertoire and structure when anticipating different types of reward," Animal Behaviour, 130, 79–96. DOI: 10.1016/j.anbehav.2017.05.025

McGrath, N., Phillips, C. J. C., Burman, O. H. P., Dwyer, C. M., & Henning, J. (2024). "Humans can identify reward-related call types of chickens," Royal Society Open Science, 11, 231284. DOI: 10.1098/rsos.231284

Morita, T., Koda, H., Okanoya, K., & Tachibana, R. O. (2021). "Measuring context dependency in birdsong using artificial neural networks," PLoS Computational Biology, 17(12), e1009707. DOI: 10.1371/journal.pcbi.1009707

Neethirajan, S. (2024a). "Decoding the language of chickens: An innovative NLP approach to enhance poultry welfare," bioRxiv. DOI: 10.1101/2024.04.29.591707

Neethirajan, Suresh. "Vocalization Patterns in Laying Hens-An Analysis of Stress-Induced Audio Responses." *bioRxiv* (2023): 2023-12. https://doi.org/10.1101/2023.12.26.573338

Nicholson, D., & Cohen, Y. (2023). "vak: A neural network framework for researchers studying animal acoustic communication," Proceedings of the 22nd Python in Science Conference (SciPy). DOI: 10.5281/zenodo.6808839

Nicholson, D. (2023). "Crowsetta: A Python tool to work with any format for annotating animal vocalizations and bioacoustics data," Journal of Open Source Software, 8(84), 5338. DOI: 10.21105/joss.05338

Niu, X., Jiang, Z., Peng, Y., Huang, S., Wang, Z., & Shi, L. (2022). "Visual cognition of birds and its underlying neural mechanism: A review," Avian Research, 13, 100023. DOI: 10.1016/j.avrs.2022.100023

Paz-y-Miño C, G., Bond, A., & Kamil, A. (2004). "Pinyon jays use transitive inference to predict social dominance," Nature, 430, 778–781. DOI: 10.1038/nature02723

Puswal, S. M., & Liang, W. (2024). "Acoustic features and morphological parameters of the domestic chickens," Poultry Science, 103(6), 103758. DOI: 10.1016/j.psj.2024.103758

Rauch, L., Schwinger, R., Wirth, M., Heinrich, R., Huseljic, D., Herde, M., Lange, J., Kahl, S., Sick, B., Tomforde, S., & Scholz, C. (2024). "BirdSet: A large-scale dataset for audio classification in avian bioacoustics," arXiv preprint, arXiv:2403.10380. DOI: 10.48550/arXiv.2403.10380

Recalde, N. M. (2023). "Pykanto: A python library to accelerate research on wild bird song," arXiv preprint. DOI: 10.1111/2041-210X.14155

Rugani, R., Fontanari, L., Simoni, E., Regolin, L., & Vallortigara, G. (2009). "Arithmetic in newborn chicks," Proceedings of the Royal Society B: Biological Sciences, 276, 2451–2460. DOI: 10.1098/rspb.2009.0044

Sainburg, T., Thielk, M., & Gentner, T. Q. (2020). "Finding, visualizing, and quantifying latent structure across diverse animal vocal repertoires," OpenReview. DOI: 10.4258/MEQ_DSSJam_

Schneider, S., Baevski, A., Collobert, R., & Auli, M. (2019). "wav2vec: Unsupervised Pre-training for Speech Recognition," arXiv preprint, arXiv:1904.05862. DOI: 10.48550/arXiv.1904.05862

Serbessa, A., Geleta, Y. G., & Terfa, I. O. (2023). "Review on diseases and health management of poultry and swine," International Journal of Avian & Wildlife Biology, 7(1), 27–38. DOI: 10.15406/ijawb.2023.07.00187

Smit, I. H., Mellbin, Y., Ask, K., te Moller, N. C. R., & Lundblad, J. (2024). "Quantifying facial expressions of the horse with optical motion capture and surface electromyography; a proof of concept," Proceedings of the 13th International Conference on Methods and Techniques in Behavioral Research, Aberdeen, Scotland. DOI: 10.6084/m9.figshare.25897855

Soster, P., Grzywalski, T., Hou, Y., Thomas, P., Dedeurwaerder, A., De Gussem, M., Tuyttens, F., Devos, P., Botteldooren, D., & Antonissen, G. (2024). "A Machine Learning Approach for Broiler Chicken Vocalization Monitoring," SSRN. DOI: 10.2139/ssrn.4999753

Stowell, D. (2023). "Computational bioacoustics with deep learning: A review and roadmap," PeerJ. DOI: 10.7717/peerj.13152

Swaminathan, B., Jagadeesh, M., & Vairavasundaram, S. (2024). "Multi-label classification for acoustic bird species detection using transfer learning approach," Ecological Informatics, 80, 102471. DOI: 10.1016/j.ecoinf.2024.102471

Thomas, P., Grzywalski, T., Hou, Y., Soster de Carvalho, P., De Gussem, M., Antonissen, G., Tuyttens, F., De Poorter, E., Devos, P., & Botteldooren, D. (2024). "Broiler chicken vocalization analysis during a medium-scale heat stress experiment," Internoise and Noisecon Congress and Conference Proceedings, INTER-NOISE24, 9537–9548. DOI: 10.3397/IN_2024_4262

Tzschentke, B., & Rumpf, M. (2024). "Acoustic communication between bird embryos - clicking sound and hatch synchronization," European Poultry Science / Archiv für Geflügelkunde, Issue 396, p. 11. DOI: 10.1399/eps.2024.396

van Merriënboer, B., Hamer, J., Dumoulin, V., Triantafillou, E., & Denton, T. (2024). "Birds, bats and beyond: Evaluating generalization in bioacoustics models," Frontiers in Bird Science. DOI: 10.3389/fbirs.2024.1369756

Vaswani, A., Shazeer, N., Parmar, N., Uszkoreit, J., Jones, L., Gomez, A. N., Kaiser, L., & Polosukhin, I. (2017). "Attention Is All You Need," arXiv preprint, arXiv:1706.03762. DOI: 10.48550/arXiv.1706.03762

Xu, R.-Y., & Chang, C.-L. (2024). "Deep Learning-Based Poultry Health Diagnosis: Detecting Abnormal Feces and Analyzing Vocalizations," 2024 10th International Conference on Applied System Innovation (ICASI), Kyoto, Japan, 55–57. DOI: 10.1109/ICASI60819.2024.10547723

Zhang, X., & Zhou, Y. (2023). "Syllable clustering analysis-based passive acoustic monitoring technology and its application in bird monitoring," Journal of Environmental Science. DOI: 10.17520/biods.2022370

Zhao, S., Cui, W., Yin, G., Wei, H., Li, J., & Bao, J. (2023). "Effects of Different Auditory Environments on Behavior, Learning Ability, and Fearfulness in 4-Week-Old Laying Hen Chicks," Animals: An Open Access Journal from MDPI, 13(19), 3022. DOI: 10.3390/ani13193022